\documentclass[%
notitlepage,
superscriptaddress,
 amsmath,amssymb,
longbibliography
]{revtex4-1}

\usepackage{graphicx}
\usepackage{dcolumn}
\usepackage{bm}
\usepackage{hyperref}
\usepackage{comment}
\usepackage[mathlines]{lineno}

\usepackage{color,hyperref}
\definecolor{darkblue}{rgb}{0.2,0.2,0.6}
\hypersetup{colorlinks,breaklinks,
            linkcolor=darkblue,urlcolor=darkblue,
            anchorcolor=darkblue,citecolor=darkblue}

\begin{document}
\preprint{APS/123-QED}

\title{Air-cushioning effect and Kelvin-Helmholtz instability before the {slamming} of a disk on water}

\author{Utkarsh Jain}
\email{u.jain@utwente.nl}
\affiliation{Physics of Fluids Group and Max Planck Center Twente for Complex Fluid Dynamics,
MESA+ Institute and J. M. Burgers Centre for Fluid Dynamics, University of Twente,
P.O. Box 217, 7500AE Enschede, The Netherlands}
\author{Ana\"{i}s Gauthier}
\affiliation{Physics of Fluids Group and Max Planck Center Twente for Complex Fluid Dynamics,
MESA+ Institute and J. M. Burgers Centre for Fluid Dynamics, University of Twente,
P.O. Box 217, 7500AE Enschede, The Netherlands}
\author{Detlef Lohse}
\affiliation{Physics of Fluids Group and Max Planck Center Twente for Complex Fluid Dynamics,
MESA+ Institute and J. M. Burgers Centre for Fluid Dynamics, University of Twente,
P.O. Box 217, 7500AE Enschede, The Netherlands}
\affiliation{Max Planck Institute for Dynamics and Self-Organization, Am Fassberg 17, 37077
G\"{o}ttingen, Germany}
\author{Devaraj van der Meer}
\affiliation{Physics of Fluids Group and Max Planck Center Twente for Complex Fluid Dynamics,
MESA+ Institute and J. M. Burgers Centre for Fluid Dynamics, University of Twente,
P.O. Box 217, 7500AE Enschede, The Netherlands}

\date{\today}

\begin{abstract}
The macroscopic dynamics of a droplet impacting a solid is crucially determined by the intricate air dynamics occurring at the vanishingly small length scale between droplet and substrate prior to direct contact. Here we investigate the inverse problem, namely the role of air for the impact of a horizontal flat disk onto a liquid surface, and find an equally significant effect. Using an in-house experimental technique, we measure the free surface deflections just before impact, with a precision of a few micrometers. Whereas stagnation pressure pushes down the surface in the center, we observe a lift-up under the edge of the disk, which sets in at a later stage, and which we show to be consistent with a Kelvin-Helmholtz instability of the water-air interface.

\end{abstract}

\nopagebreak

\maketitle

\nopagebreak

Spectacular phenomena occur in nature when a liquid slams into a solid, observed when, e.g., ocean waves crash against a harbour quay \cite{peregrinereview}, a stone lands in a lake \cite{truscottreview}, or seabirds catch their prey \cite{Chang12006}. Similar phenomena can be seen during the operation of ocean vessels \cite{abrate2011hull} or the landing of a space vehicle on the ocean \cite{SEDDON20061045}. The traditional understanding of these impact phenomena is achieved in the context of macroscopic potential flow analysis, that involves the transfer of momentum from the solid to the liquid phase, known as the added mass effect \cite{batchelor1967introduction, bagnold1939interim,glasheen1996vertical,chuang1966experiments,peregrine1996effect,korobkinreview,wagner1932stoss,howison_ockendon_wilson_1991,ERMANYUK2005345}.

More recently, it is becoming clear that this macroscopic picture is insufficient to fully understand the underlying physics. By acting mostly on a microscopic scale, the intermediate air phase causes effects that may propagate towards the largest scales, i.e., determining whether or not a splash occurs, which can correspondingly be controlled by changing the air pressure \cite{KolinskiBrenner2012,MandreBrenner2009,RibouxGordillo2014}, or the wetting properties of the substrate \cite{RibouxGordillo2014} during droplet impact. Vice versa, on small scales, the remnant of the air layer can be the entrapment of a tiny air bubble at impact, whose size is set by force balances on a much larger scale \cite{bouwhuis2012a, tran2013jfm}.

In all the above cases however, the phenomena arise from the existence of a small length scale in lateral direction, namely the impact point. What happens when the impact {region} is laterally more extended, i.e., for the impact of a flat disk onto a liquid surface? In this paper we will address this question, and show how the water surface is affected in the very last stages before impact, when the air is squeezed out from the gap between the liquid and the disk at increasing speeds. We will experimentally demonstrate that the resulting surface deformation is quite different for this case, supporting and substantiating prior speculations based on qualitative grounds \cite{verhagen1967impact, watanabe1988observation, oh2009, hicks2012air}, and on numerical \cite{Barjasteh18, peters2013splash} and theoretical arguments \cite{smith2003air, hicks2012air, wilson1991mathematical, bouwhuis2015initial}. Here, we combine high-speed imaging with a novel deflectometric technique which allows us to measure the {vertical} movements of the water surface just before the impact of a flat disk with micron precision, and succeed for the first time to experimentally characterize the motion of the free surface.

\begin{figure}
\includegraphics[width=.7\linewidth]{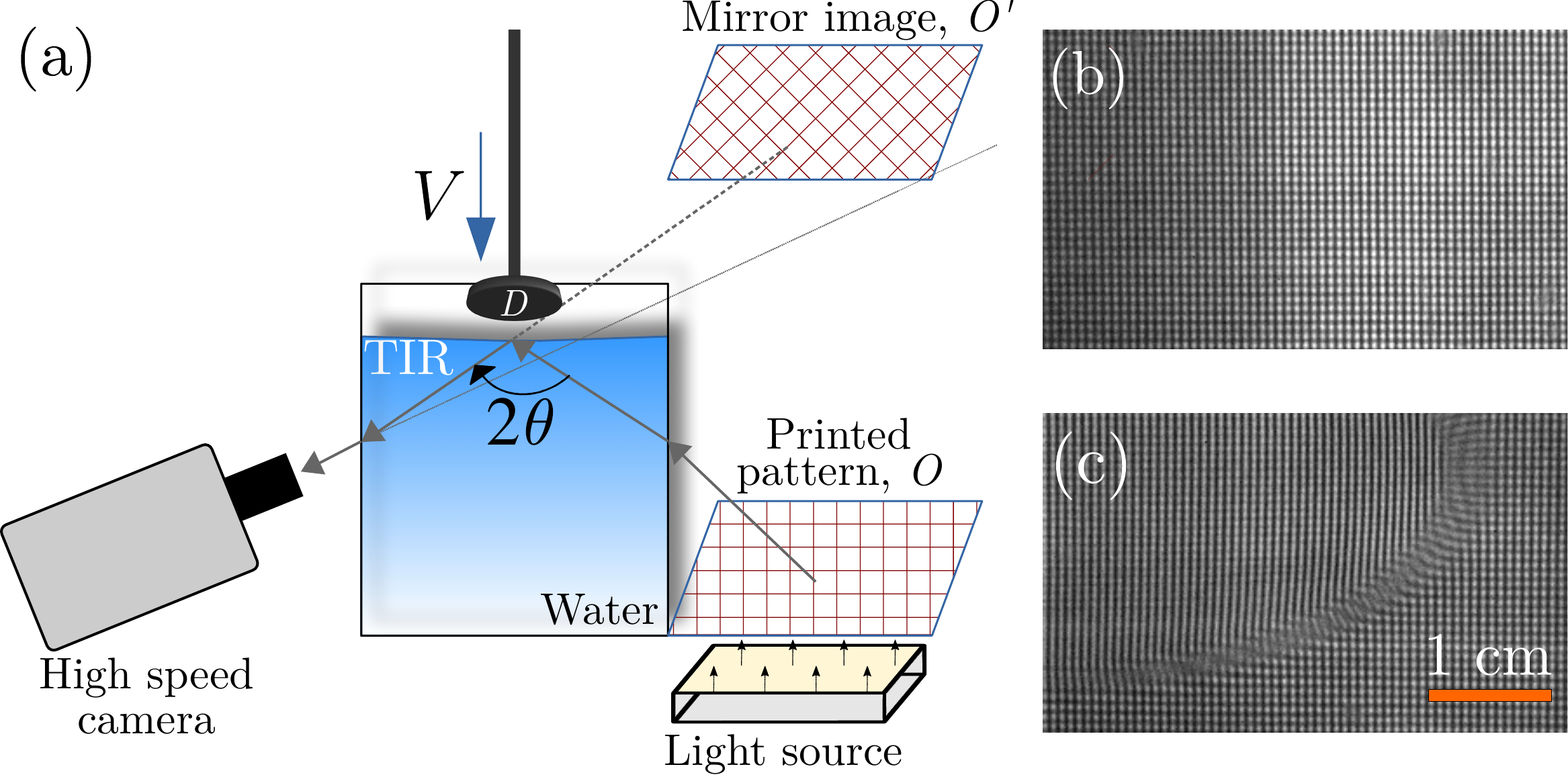}
\caption{(a) Experimental setup: a light source  illuminates a reference pattern $O$ with printed lines,
which reflects from the water-air interface into a high-speed camera. At high incident angle $\theta$, the water-air interface acts as a mirror, and the camera records the mirror image $O'$, as indicated by the light rays (in gray). (b) Image of a stationary air-water interface reflecting the reference pattern. (c) Deformed air-water interface reflecting a distorted image of the reference pattern.}\label{fig:KHsetup}
\end{figure}

\begin{figure*}
\centering
\includegraphics[width=.99\linewidth]{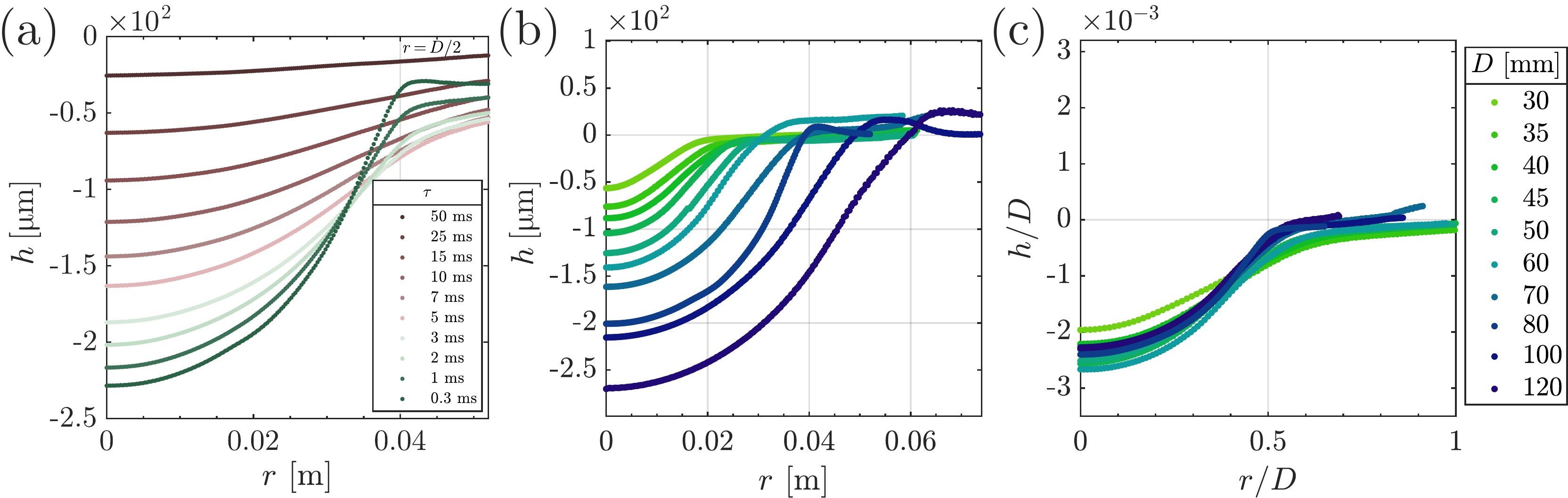}
\caption{(a) Free surface profiles $h(r)$, azimuthally averaged about the disk center, shown at different times $\tau$ before impact (occurring at $\tau=0$) from an experiment with an 8 cm wide disk (edge shown by vertical gray line) approaching the free surface at 1 m/s. (b) Free surface 
$\tau = 0.033$ ms before impact for a range of disk diameters as indicated in the legend (shared with panel (c)). (c) Non-dimensionalised water-surface profiles at non-dimensionalised time $\tau V/D = 0.01$ before impact for the same 
disk sizes. See supplementary material \cite{supplementary} for videos of the time evolution of the surface profile for $D = $ 50 and 80 mm.}\label{fig:surfaceprofiles}\end{figure*}

Our setup (figure \ref{fig:KHsetup}) consists of a flat steel disk of diameter $D$ (3--12 cm), which impacts on a quiescent demineralised water bath with a controlled speed $V = 1$ m/s. The Weber number of impact $\text{We} = \rho_{\text{water}} V^2 D / \sigma_{\text{water-air}} $ {typically varies from {400 to 2500}}. To visualise the water surface hidden under the disk, we use total internal reflection, {where the liquid-air interface acts as a mirror, and is monitored from below. A high-speed camera and an illuminated reference pattern are placed on either sides of the tank (figure \ref{fig:KHsetup}(a)), and positioned {such} that an incoming light ray from the pattern {is} fully reflected at the interface. While} a stationary free surface simply reflects back the image of a reference pattern (figure \ref{fig:KHsetup}(b)), the deformed free surface reflects a distorted image (figure \ref{fig:KHsetup}(c) {and supplementary movie} \cite{supplementary}). Using the ray optics in the setup \cite{tirdmanuscript, ujthesis}, the liquid surface is reconstructed with a method inspired by (refraction based) synthetic Schlieren technique \cite{moisy2009synthetic, wildeman2018real}. The displacement fields from a movie \cite{supplementary} of the deforming reference field are collected, and used for a full spatial reconstruction of the free surface with an unprecedented precision of a few micrometres.

The typical deformation of the free surface just before the impact is shown in figure \ref{fig:surfaceprofiles}(a) and in the supplementary movie 2 \cite{supplementary}. The azimuthally averaged profile $h(r,\tau)$ is plotted as a function of the radial coordinate $r$, (with $r=0$ the point just below the impacting disk's centre) for varying times before impact $\tau$. The initially quiescent liquid-air interface is deformed by the air flow more than 50 ms before the disk makes direct contact with the bath. The free surface gets increasingly pushed down with time, and forms a 200 $\mu$m deep cavity just before impact. In addition, an opposite effect is observed in the last 5 ms: the free surface is lifted up by $\sim$10 $\mu$m close to the disk edge (visible in figure \ref{fig:surfaceprofiles}(a)). Both effects are seen for a range of disk sizes as presented in figure \ref{fig:surfaceprofiles}(b), where we plot the free surface profile just before impact (at $\tau = 0.033$ ms). The liftup of the free surface is more pronounced for larger disks. As a result, the solid surface of the impacting disk makes its first contact with the water surface along its periphery, thereby entrapping an air bubble upon impact. This local pull and the central push on the free surface were both already reported in a qualitative manner by Verhagen \cite{verhagen1967impact}, but thus far could not be measured.

Both these effects are a consequence of the squeezed air flow, escaping from below the disk. The average velocity profile of air in the gap between the undeformed liquid surface and disk is estimated using a depth-integrated continuity equation (more details in supplementary material \cite{supplementary} and refs \cite{ishizawa1966unsteady, jackson1963study})
\begin{align}\label{eqn:airflow}
\mathbf{V}_{r,\text{gas}} = \frac{r}{2 \tau} \hat{e}_r.
\end{align}
Note that the Reynolds number in the air {gap of width $d=V\tau$ under the disk edge}, $Re = \rho_{\text{gas}}V_{R,\text{gas}} d/\mu_{\text{gas}} = \rho_{\text{gas}} V D / 2\mu_{\text{gas}}$, is much larger than unity for all  $D$ and $\tau < 50$ ms, so that viscous effects in the air layer may be ignored \cite{peters2013splash, jackson1963study}. {Additionally, an alternative definition for reduced Reynolds number suited particularly for lubrication flows, defined as $\delta^2 Re$, where $\delta$ is the aspect ratio of the air film, would also yield estimations much larger than unity. For these reasons, we emphasise that lubrication effects are absent in the present problem, and the air flow can be safely treated within the framework of inviscid flow. We also remark that Mach numbers of the escaping air reach values above 0.3 only in the very last instants ($\tau \approx 0.39$ ms for the example shown in figure \ref{fig:surfaceprofiles}(a)) using the definition in equation \eqref{eqn:airflow}. Thus, any compressibility effects in air may be neglected for the range of free surface dynamics we study.} Equation \eqref{eqn:airflow} {then} has an important consequence in that there is a stagnation point at $r=0$. It causes a pressure build-up in the squeeze-layer of air, an effect evidenced {both} theoretically {and experimentally} for squeeze flows between a solid and a fluid phase \cite{bouwhuis2012a, bouwhuis2015initial, wilson1991mathematical, hicks2012air, hicks2013liquid, KolinskiBrenner2012, mani_mandre_brenner_2010, klaseboer2014universal, hendrix2016a, CHERDANTSEV2021110375}. Previously estimated thickness of the trapped air layer after impact from experiments \cite{mayer2018flat} indicate a surface cavity of a size that is consistent with our experiments.

Another consequence of equation \eqref{eqn:airflow} is that for a disk of diameter $D$, the escaping air has its largest velocity under the disk edge at $r=D/2$, just before being pushed out into the ambient air. In keeping with the Bernoulli principle, the increasing flow speed of air under the disk creates a region of low pressure where its velocity is the highest, {i.e., under the disk edge, where it generates a pull on the water surface, causing it to be drawn into the gap.}

\begin{figure}
    \centering
    \includegraphics[width=.6\linewidth]{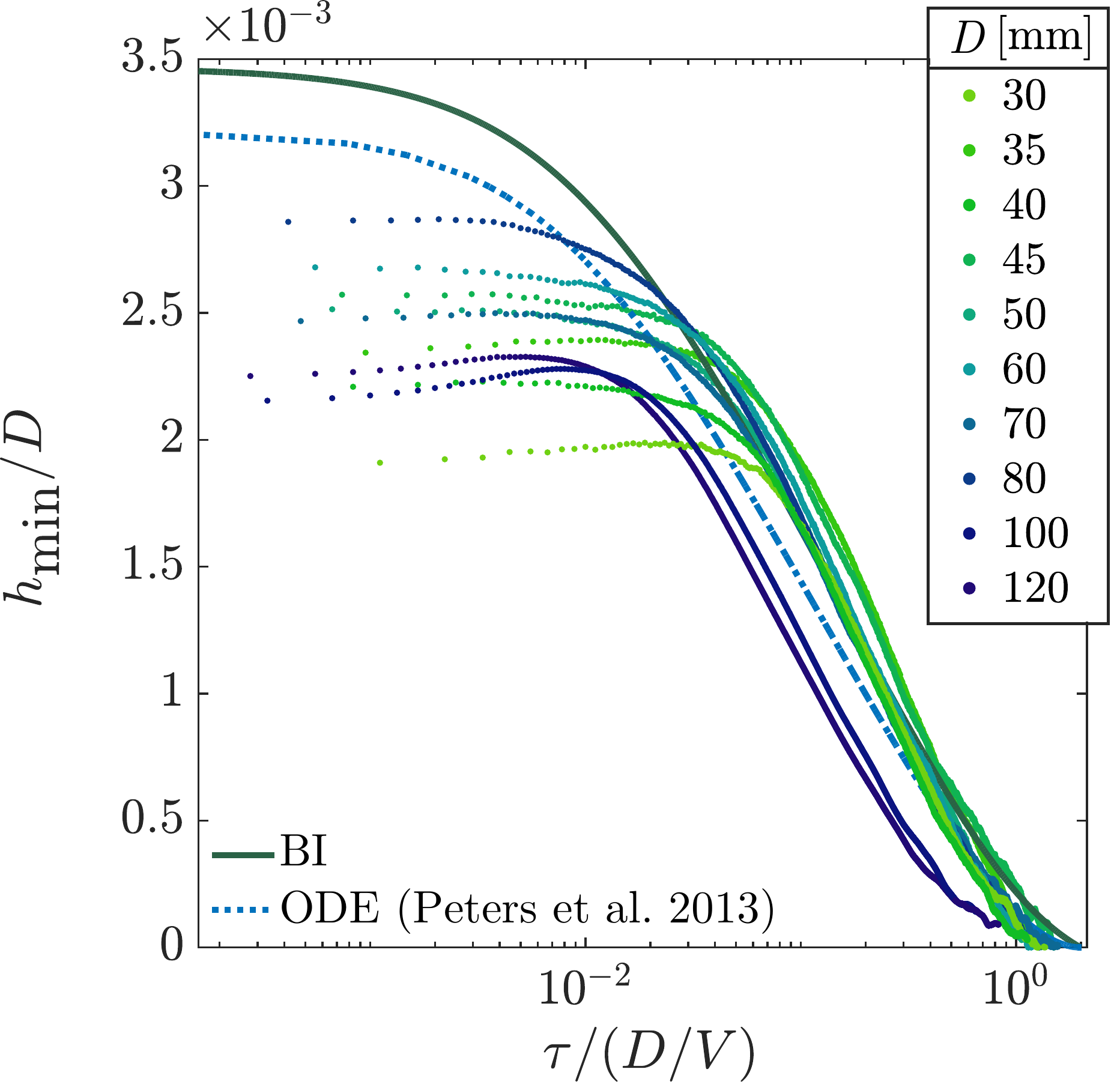}
    \caption{{Non-dimensionalised cavity depth $h_{\text{min}}/D$ (measured at $r=0$) as a function of the non-dimensionalised time before impact $\tau V /D$, with $D$ varying between 30 and 120 mm. The solid and dotted lines are the two-fluid boundary integral simulation and analytical calculations results respectively from Peters et al.~\cite{peters2013splash}.}}
    \label{fig:minsurfplot}
\end{figure}

The quantitative analysis of our experiments gives more insight into the dynamics of the free surface deformation. Since $\mathit{Re} \gg 1$ in the air layer, the formation of pressure gradients in this region is an inertially driven process, with the relevant length and time scales being $D$ and $D/V$. This is made evident when considering the time evolution of the central cavity, as shown in figure \ref{fig:surfaceprofiles}(c). All the free surface profiles taken at the same non-dimensionalised time before impact $\tau V /D$ overlap in a reasonable way when $h$ and $r$ are rescaled by $D$. Further insight into the mechanism that drives the growth of the depth $h_{\text{min}}$ of the central cavity is obtained by looking at its time-evolution, shown in figure \ref{fig:minsurfplot}. Experimental data for varying disk sizes $D$ are collected, non-dimensionalised by $D$ and $D/V$ respectively, and compared to results from two-fluid boundary-integral simulations from Peters et al. \cite{peters2013splash} (green line, figure \ref{fig:minsurfplot}). We also show (as a blue dotted line) the result of an analytical calculation by the same authors,
which simplifies the problem by neglecting the radial variation of $h$ with $r$ ({while} ignoring the cavity{'s closure at $r=R$}).
We find good agreement between our experimental measurements and the previous results \cite{peters2013splash}, especially for the smaller disk sizes ($D < 80$ mm), and $\tau V/D \gtrsim 2 \cdot 10 ^ {-2}$. In particular, the boundary integral simulations \cite{peters2013splash}, which provide a more complete picture, show an excellent match with our experimental data, without a single free parameter. The good agreement between experiments, model and simulations regarding $h_{\text{min}}$ persists into the later regime when its growth rate saturates. The final value of $h_{\text{min}}/D$ {just before impact} is however 20\% to 30\% smaller in experiments.

\begin{figure}
    \centering
    \includegraphics[width=.99\linewidth]{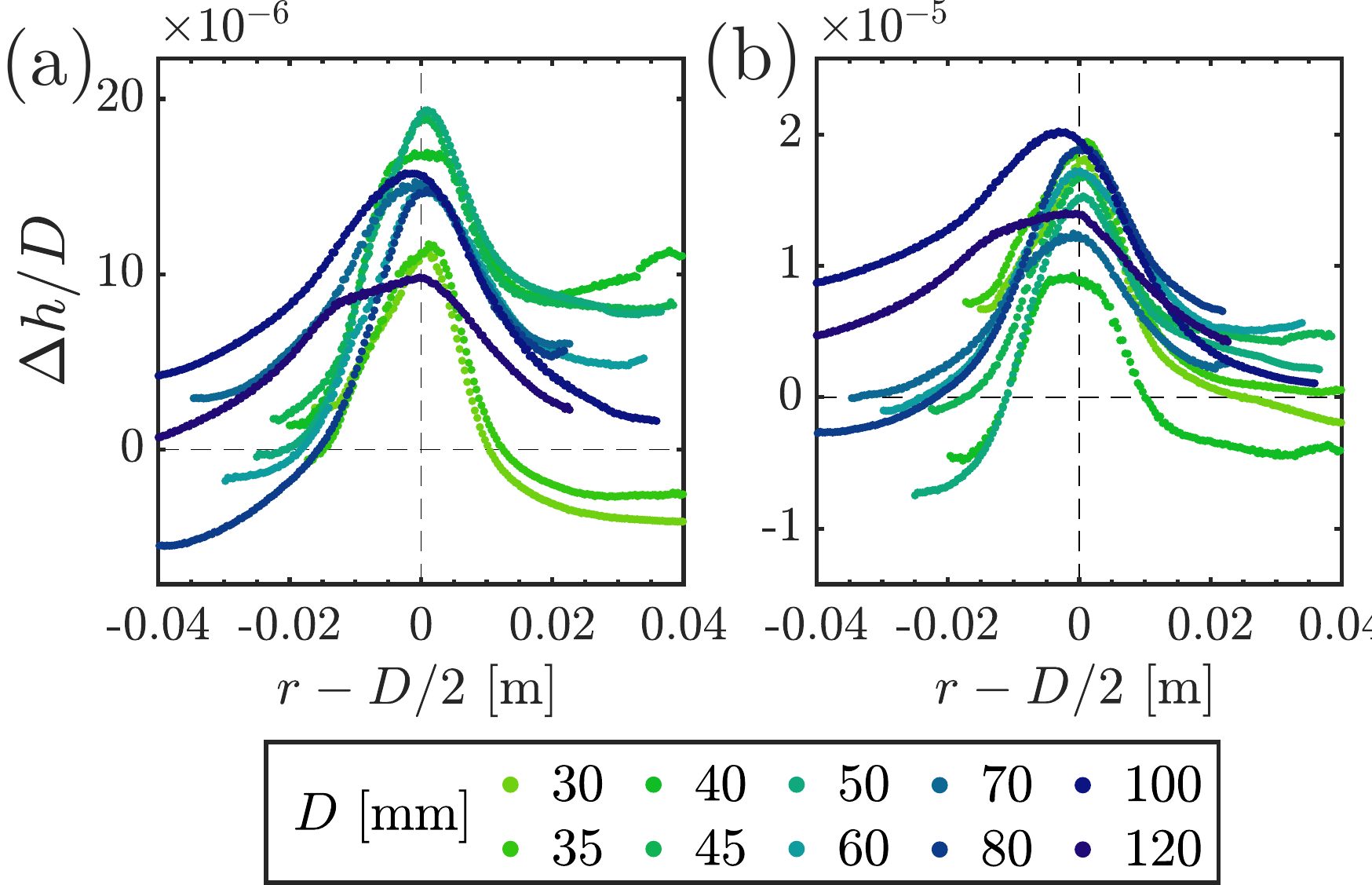}
    \caption{The instantaneous, dimensionless {vertical} deflection {velocity} $\Delta h/D = \left( \partial h / \partial t \right) \left( \Delta \tau / D \right)$ of the surface profile at dimensionless times (a) $\tau V /D =$ 0.005 and (b) $\tau V /D =$  0.003 before impact. $\Delta \tau =1/30000 \text{ s}$ is the time between two successive images in an experiment. The profiles are plotted with radial coordinates $r$ shifted by $D/2$ such that they are all centred under the disk edge.}
    \label{fig:shiftedsurface}
\end{figure}

To better understand this, we now focus on the liquid suction phenomenon, which has been reported in boundary-integral simulations \cite{peters2013splash, Barjasteh18}, but never measured experimentally. As visible in figure \ref{fig:surfaceprofiles}(a), the growth of the bulge (of height $h_{\text{max}}$) starts at $\tau \approx$ 5 ms, i.e., much later than the cavity, which was already visible at $\tau = $ 50 ms. It is an order of magnitude smaller than the latter. To characterise its growth, we choose to focus on the instantaneous velocity $\partial h(r,{t})/\partial t$ with which the free surface deflects. In figure \ref{fig:shiftedsurface}(a) and (b) the velocities of the upwards moving free surface are shown at two different dimensionless times $\tau V /D = 0.005$ and $0.003$ before impact. For varying disk size $D$ the profiles are plotted with shifted radial coordinate $r - D/2$ so that they all overlap at the disk edge. Figure \ref{fig:shiftedsurface} reveals not only that the suction clearly acts the strongest just under the disk edge, but also that it acts over a consistent length scale $\lambda \lesssim 2 \text{ cm}$, that appears to be independent of the disk diameter $D$. Thus, contrary to the cavity formation, the segment of the free surface which is drawn upwards is not affected by the inertial length scale set by the disk size $D$. {Note that such a consistent length scale over which the suction is seen to act in figure \ref{fig:shiftedsurface} is remarkably robust despite the control parameter $D$ being varied over a factor of four. This is made more evident from the results obtained with smaller disks with radii of up to 1.5 cm, where the unstable wavelength itself is larger than the radius of the disk whose approach causes the interface to be destabilised. In comparison with a disk of radius 6 cm, the length scale being sucked towards it with highest velocity is less than $1/3$ of the disk radius.}

Regarding the localization of the bulge at the rim, a similar observation was made by Oh et al.~\cite{oh2009} with a large rectangular impactor (of width 30 cm) for which they interestingly do not observe a single bump, but spatially periodic ripples along the free surface, localized under the impactor's edge.

The above observations suggest the role of a regular shear instability at the sharp water-air interface behind initiating the elevation of the liquid surface well before impact. Indeed, such a mechanism dictates that the destabilised wavelengths at the liquid surface lie within a close range, within limits determined by the competition between gravity and surface tension. The balance between these two restoring forces yields the most-unstable wavelength $\lambda^{\text{marg}}$ at the onset of instability. We expect $\lambda^{\text{marg}}$ to be independent of $D$, as the inertial length scale does not affect the properties of the water-air interface. \cite{hsieh1994wave,chandrasekhar1981hydrodynamic}

To verify this hypothesis, a linearised Kelvin-Helmholtz analysis was performed assuming normal-mode perturbations along the water-air interface, subjected to a shear air flow in the inviscid approximation.
Here, the configuration is slightly more involved than the classical one: the upper fluid layer (the escaping air layer) has a finite thickness $d ({\equiv} V \tau)$, which affects the condition of marginal stability of the interface perturbation (see supplementary materials \cite{supplementary}).
In particular, a decreasing $d$ can modify the marginally unstable solution if $d^{\text{marg}} \lesssim 7$ mm. However, our experiments are wholly performed for disk sizes $D$ and velocities $V$ such that $d^{\text{marg}} \gtrsim 6$mm, so that the solutions for an interface separating two infinitely deep fluid half spaces are expected to work within reasonable variance. With this assumption, the marginally unstable wavelength $\lambda^{\text{marg}}$ is
\cite{hsieh1994wave} \begin{equation}\label{eqn:lambdamarg}
    \lambda^{\text{marg}} = 2\pi \Big[ \frac{g}{\sigma} (\rho_{\text{w}} - \rho_{\text{a}}) \Big]^{-1/2} \approx 1.7 \text{ cm},
\end{equation}
where $g$ is the gravitational acceleration, $\sigma$ the surface tension, and $\rho_{\text{w}}$, $\rho_{\text{a}}$ the fluid densities of water and air, respectively. The marginally unstable wave length found by the calculation above is close to that shown in figure \ref{fig:shiftedsurface}, and independent of $D$ and $V$.

\begin{figure*}
    \centering
    \includegraphics[width=.99\linewidth]{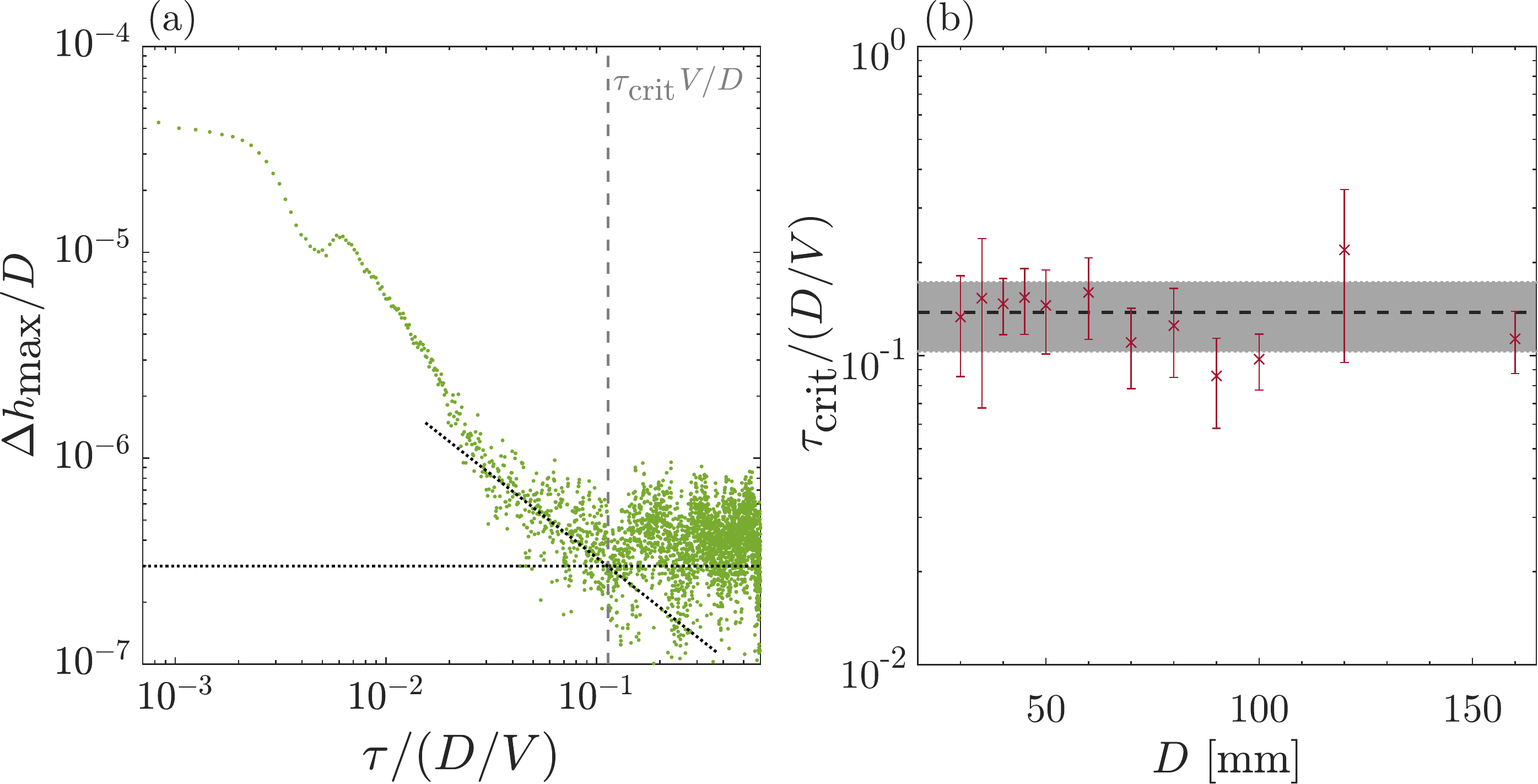}
    \caption{{(a) Time evolution before impact of the maxima of dimensionless free surface deflection velocities $\Delta h/D = \left( \partial h / \partial t \right) \Delta \tau / D$ for a disk of width 16 cm. The free surface starts to accelerate towards the disk close to $\tau _{\text{crit}} V /D \approx 0.11$ (grey dashed-dotted line), which is estimated by tracing the growth rate (using dotted lines). (b) $\tau _{\text{crit}} V /D $ measured over several $D$ are shown. The dashed line shows the mean of the measurements, while the grey box shows the range of standard deviations of the measurements.}}
    \label{fig:suctionvelwithtime}
\end{figure*}

Another key finding from the spatial analysis of the dispersion relation is that the minimum air velocity $V_{\text{air}}^{\text{min}}$ for the water-air interface to become unstable also deviates from an infinitely-deep fluids configuration when $d^{\text{marg}} \lesssim 7$ mm. Without any finite-depth effects in either fluid, the usual calculation yields \cite{hsieh1994wave} \begin{equation}\label{eqn:critvel}
    V_{\text{air}}^{\text{min}} = \sqrt{ \frac{2 (\rho_{\text{w}} + \rho_{\text{a}})}{\rho_{\text{w}}\rho_{\text{a}}}   \Big\{ \sigma g (\rho_{\text{w}} -
    \rho_{\text{a}})   \Big\}^{1/2} } \approx 6.58 \text{ m/s},
\end{equation}
with the connotation that linear Kelvin-Helmholtz analysis at the water-air interface is known for over-predicting critical velocity \cite{paquier2016generation, funada2001viscous}. This makes it difficult to directly compare equation \eqref{eqn:critvel} with our experiments. However, {since Kelvin-Helmholtz instability is supercritical in nature, occurring once a driving parameter has crossed a critical threshold, an equivalent measure of $V_{\text{air}}^{\text{min}}$ in our experiment may be obtained from the non-dimensionalised time $\tau V /D $ at which the interface's suction starts to accelerate. One instance of estimating the critical time before impact $\tau _{\text{crit}} V /D $ is shown in figure \ref{fig:suctionvelwithtime}(a). The fact that $\tau _{\text{crit}} V /D $ measured in experiments over several $D$ is constant, as shown in figure \ref{fig:suctionvelwithtime}(b), clarifies that the instability threshold {$V_{\text{R,gas}} ^\text{min} = R/2 \tau_{\text{crit}} \approx 2.27 $ m/s} in our experiment is indeed independent of the disk size.} {Note that it may not be immediately obvious that such a frozen-time stability analysis should be applicable in the present scenario. Indeed, a rapidly accelerating base flow in general may demand the temporal aspects in the dispersion relation to be analysed. However, since the onset of instability in our experiment occurs in conditions when the base flow (of air) is always of the order of few metres per second, such considerations have limited bearing on the instability criterion in the present experiment.}


To conclude, we report the first experimental measurements of the water-interface deflections caused by the air cushioning effect prior to a solid-liquid slamming event. This is achieved using a novel TIR based measurement technique that is able to resolve micron-scale {vertical} movements of the reflecting water-air interface.
In the air cushion under a slamming disk, a peculiarly strong air flow is set up under the edge, which locally results in an upwards suction of the liquid surface. The suction is shown to act over a unique length scale that remarkably is independent of the inertial parameters, indicating the action of a regular interfacial instability. The experimental data show that the suction is initiated when the air velocity under the disk edge grows beyond a critical value, as typical for a Kelvin-Helmholtz instability.

The interplay of the inertial deflection of the water surface away from the disk and the localised suction towards it invariably creates conditions for an air film to be trapped immediately after impact. This has significant implications on how the slamming pressures are distributed across the impacting disk's surface both in space and in time \cite{ujthesis}. But more importantly, both the inertial and instability mechanisms will also operate on the liquid surface at larger scales, such as in hull slamming \cite{abrate2011hull, kapsenberg2011slamming}, sloshing of liquid in cargo transport \cite{dias2018slamming}, or even in water landing of space-vehicles \cite{SEDDON20061045}, where much more violent water-slamming events are encountered, accompanied by stronger gas flows, thereby leading to even more significant modification of the pressure loads. At these large scales, the Kelvin-Helmholtz instability of the liquid interface may be utilized by modifying the base flow of air: the most unstable modes can either be selectively amplified, or indeed be suppressed by invoking neighbouring parasitic modes. In this manner, by exploiting the instability one may sculpt the shape of the liquid interface and thus optimize the distribution of the impact loads on the solid target.

\section{{Acknowledgements}}
Many thanks to G-W.H. Bruggert and B. Benschop for technical support. We also express thanks to I.R. Peters and A. Prosperetti for helpful diskussions. We acknowledge funding from SLING (project number P14-10.1), which is partly financed by the Netherlands Organisation for Scientific Research (NWO).

\bibliography{apssamp}


\end{document}